# Redrafting Requirements Modeling Using *a Single* Multilevel Diagram

Sabah Al-Fedaghi[*]

*Computer Engineering Department*
*Kuwait University*
*Kuwait*

salfedaghi@yahoo.com, sabah.alfedaghi@ku.edu.kw

*Abstract* – The complexity of software-based systems has increased significantly, especially with regards to capturing requirements along with dependencies among requirements. A conceptual model is a way of thinking about and making sense of the real world's complexities. In this paper, we focused on two approaches in this context: (a) multiple models applied to the same system with simultaneous usage of dissimilar notations vs. (b) a single model that utilizes a single framework of notations. In the first approach, inconsistencies arise among models that require a great deal of painstaking discipline and coordination between them. The multiple-model notion is based on the claim that it is not possible to present all application views in a single representation, so diverse models are used, with each model representing a different view. This article advocates a second approach that utilizes a single model with multilevel (static/dynamic and behavioral) specification. To substantiate this approach's feasibility, we embrace the *occurrence-only model*, which comprises (a) Stoic ontology, (b) thinging machine (TM) language and (c) Lupascian logic. In this paper, we focus on TM modeling as the mechanism of single-model building. We claim that a TM can be a unifying diagrammatic language for virtually all current modeling languages. To demonstrate such a claim, we redraft almost all the diagrammatic representations in *The Handbook of Requirements Modeling of the International Requirements Engineering Board.* **This redrafting includes context, class, activity, use case, data flow and state diagrams. The results seem to indicate that there are no difficulties in representing all views in TM.**

## I. Introduction

The real world's increasing complexity suggests the need to reconsider how to carry out conceptual modeling [1]. A conceptual model is "a way of thinking about and making sense of the complexities of the real world" [2]. In this situation, maintaining consistency is a crucial aspect of model-driven software engineering [3].

Additionally, the rate at which software-based systems' complexity grows has increased significantly, especially regarding capturing requirements along with dependencies among requirements [4]. Precise requirements analysis and capture are of great significance in identifying user needs in the early stages of the development process.

------------------------------------------
*Retired June 2021, seconded fall semester 2021/2022

Errors that resulted from improperly defined requirements lead to additional effort in respecification, redesign, redevelopment and retesting. In this case, costs increase tremendously.

These facts clearly show how important the precise requirements analysis and capture are [5]. It is essential that the requirements are elaborated to such an extent that they are easy to read and unambiguous but still offer the developers enough freedom in their implementation [5].

System development involves a multi-stakeholder (including software engineers) effort that requires a *shared understanding* of the system requirements [6]. Problems with requirements arise when different stakeholders envisage different solutions for a given problem, with the consequence that they specify different, conflicting requirements for that problem [6].

Requirements modeling will help enhance shared understanding among stakeholders. Modeling is a particular way of ratifying explicit requirements though abstract representation of the part of reality to be created. Models help infer coarse concepts and increase the possibility of interpreting them correctly, thus contributing to the creation of proper shared understanding [7]. To specify such a model, a diagrammatic language is created to support the requirements' elicitation, analysis, and documentation process. It is a language designed specifically for easy consumption by executive, business and technical stakeholders [8].

### A. Problem

When developing many models to treat the same system with the simultaneous usage of dissimilar notations, we face genuine difficulties in keeping various system accounts consistent with each other. Other difficulties include traceability, process mapping, changeability, completeness, over-specification, conformity and documentation [9]. They necessitate a great deal of painstaking discipline and synchronization between the models. Such a multiple-model approach is based on the claim that it is not possible to present all application views in a single model, so diverse representations are used, with each model representing a different view. According to [6], "If multiple models are used to describe the requirements, it is important to keep these models consistent with each other. This requires a lot of discipline and coordination between the models." An in-depth study of UML [10] showed that the official definition of UML 2 includes inconsistent UML models.

Nevertheless, the current prevailing line of software system development promotes the notion of unification, as in *Everything is an object* [11], and unified multiple modeling. Unification refers to a unifying notation that incorporates the best of a number of other notations as well as current best practice in one generally applicable notation [12].

An alternative approach is the use of an integrative approach of a *single* model that enables requirements specification in a uniform diagrammatic language. Several attempts have been made in this direction. In this paper, we use a model called a thinging machine (TM) modeling [13–14]. For brevity, we will focus on modeling *behavior* or the change of the system aspects over time, as we will illustrate next.

### B. Motivation: Behavior Models

We specifically divide the conceptualization process into two distinct phases: a static portrait of *potentialities* and dynamic representation of *actualities*. The potentiality refers to the possibility of coming into existence as an event. The terms here are based on the ontological model, called the occurrence-only model, presented in [15], and will be reviewed briefly later in this paper. Potentiality is a basic philosophical notion with many related notions (e.g., remoteness and papery; e.g., semen, embryo, and baby [16]). In this context, we will try not to become involved in deep philosophical aspects at this phase of our research. The material in this paper is oriented toward minimizing these aspects and delaying further exploration of future research.

We use the notion of potentiality in the Stoic sense as *subsistence* not *existence* of things (see [14]) and utilize it in a static timeless level of modeling. For example, we tie this *stacity* with the notion of *state* (as in state machine) to support our claim that *state diagrams* are not exactly a *dynamic* description. In this line of thinking, if we know an object's state, we still need to know about its timed version. *A person is in love* is a (*static*) *state* condition, but we still, at the dynamic level, have to supplement the statement with an *instance* of this person (e.g., William Shakespeare time: 1582–1616). Further, we need to know when he/she fell in love (e.g., time: age 20 and up) to transform the *static* statement to a *dynamic* description. This claim implies that such (software) diagrams as UML state, activity and sequence diagrams fall short of being behavior models. In two-level modeling, the static and dynamic descriptions are separated in timeless and time-infested levels, as illustrated next.

According to [17], UML has three ways to specify *behaviors*. Each emphasizes different aspects: *activities* for inputs and outputs between actions and their time ordering, *state machines* for reacting to notification of external events and *interactions* for messages between objects. Ref. [17] explained that "the real-world implications of anything said in behavior languages are what occurs when behaviors actually happen." For example, a factory operation for changing an *object's color* happens many times every day at many factories, each involving a different object, different colors and so on. Each time the behavior happens is a separate behavior occurrence. Accordingly, it seems that [17] argue that behavior is specified in terms of *types* that are instantiated repeatedly. The semantics of behavior languages specify which *occurrences* (Fig. 1 right) cannot be expressed completely (Fig. 1 left). Additionally, [17] proposed specifying the successions between behavior events, as Fig. 2 shows.

The TM modeling, as will be demonstrated in this paper, fills the diagrammatical gap between the left and right diagrams of Fig. 1. TM treats activities, objects and states in a uniform way as *events*. We show that a TM provides an alternative ontological foundation and modeling language for most current approaches to modeling behavior.

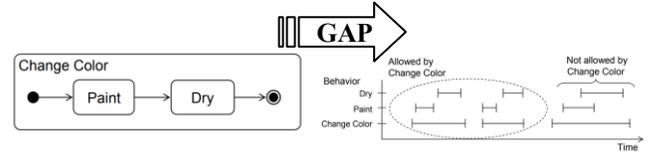

**Fig. 1 (Left) behaviour, (right) occurrences (from [17])**

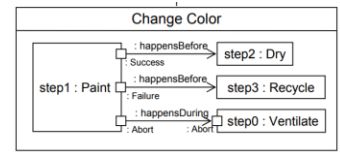

**Fig. 2 Successions between behaviour events [17]**

### C. Two-level Modeling

This section illustrates the notion of *vertical* representations over a single model.

Instead of the common approach of separate diagrammatic representation of static and dynamic features *(e.g., class vs. state diagram), a TM language assembles a model that* has vertically dynamic representation over static representation. Staticity refers to *timelessness*. The static TM model is built from subsisting *regions* with a logical order imposed by *potential* flows and triggering. The static model comprises fixed parts, and it simply *subsists*, e.g., "the flow of traffic depends on cars [and flow] without being anything but the cars" [18]. Traffic is not itself a solid body, but it is nonetheless real because it depends on cars for its *subsistence*; this subsistence is captured in the static region. Cars are entities that exist as physical things. Traffic is a process that subsists as a *region* of the existing traffic. If there are no cars; still, traffic subsists as a *potential* thing. Suppose that the traffic "disappears" even though there are cars and roads (e.g., due to COVID-19). The absence of traffic may mean,

(a) There is such a thing called traffic, but it currently does not occur (stopped).

(b) There is no such thing called traffic, just as there is no life after death and no *round square* even though there are squares and circles.

Clearly, (a) is the plausible answer. If there is a thing called traffic that is not currently actualized (exists), then *where is it*? The Stoic answer is *it subsists*.

To make the notion of subsistence clearer to computer scientists, consider the process of addition instead of traffic, expressed by the code *ADD A,B*. The instruction *ADD A,B* is of course just a code for the process of addition. The process of addition subsists in *ADD A,B*, and it comes into existence when actualized in the execution of *ADD A,B*. When it is not executed, the addition process is subsisting in *ADD A,B*. We categorize *ADD A,B* as the *static* description of addition, and when it is executed, it is realized *dynamically*, e.g., fetch operand A in ALU, fetch operand B into ALU. So, the code *ADD A,B* is not addition; it is where addition subsists. When the addition does not exist, i.e., in the execution mode, it is not in the "nonexistence" mode because it is still there in reality waiting for execution. On the other hand, *ADD A,B does not exist* signifies its disappearance or vanishing, e.g., it was erased or never written as a code. That is why the Stoics believed the meaning (*Lekta*) subsisted in linguistic expressions. Therefore, the meaning of *ADD A,B* is certainly not the linguistic symbols. It may be described as what subsists inside *ADD A,B* that can actualize in reality.

Subsistence is as real as existence. A stoppage of a process does not mean its destruction, just as the stoppage of an engine does not mean its disappearance. The engine is still there, but it is no longer running. The existence and not existence of *running* (working) depend on the engine because *running* cannot exist by itself without an engine. In this case, if *running* does not exist, it *subsists*, ready to exist again. The subsistence of running is not its *destruction* as in the *living* of humans and animals vs. *non-living* (death) that will not occur (to the living) again (potentiality) when it stops. Similarly, the stoppage of traffic means that cars and roads are *still there*, but the traffic process is still in reality as a subsisting phenomenon ready to exist again. In this case, according to the Stoics, reality can be described in the form of alternating between existence and subsistence.

Here, we see that *potentiality* is a weaker feature than *possibility*. Potentiality of existence implies a thing exists by itself or inside an existing thing. Possibility implies a thing exists by itself. Therefore, cars and roads are possible, whereas traffic is a potential (needs cars and road to exist). Because traffic's existence is based on the existence of "cars and road" and the description of traffic subsistence is a *region* that includes cars and road, the whole region is a potentiality that includes the stronger feature of possibility.

Fig. 3 shows existing traffic as a repeated *event* of car flow. Fig. 4 shows the static traffic as a *region* of process (event). The subsistence of the static traffic becomes an existing event. The region of traffic subsists as a potentiality. The notion in these two figures will be discussed in the second section of the paper.

Building a *dynamic* model involves representing *existing things* (e.g., physical things, earthquakes, yards, outer space). Existence here evolves from potentiality of the static diagram. This two-level modeling is based on the Stoic two-level *being* that adopts two kinds of reality: existence and subsistence. Such a two-level model is missing from most current modeling approaches.

**Example:** Consider the state diagram shown in Fig. 5 [19], which models a turnstile used to control access to subways. Depositing a coin in a slot of the turnstile unlocks the arms, allowing a single customer to push through. After the customer passes through, the arms are locked again until another coin is inserted.

According to [20], "Understanding *state* is fundamental to successful modeling. Everything we need to know and everything we want to do can be expressed in terms of the states of the system under control because ultimately those are the things we wish to control." Nevertheless, the state (in software modeling) is typically defined in an informal way. The OMG [21] "document Semantics of UML State Machines" mentions the term "state" 2,462 times but does not define it.

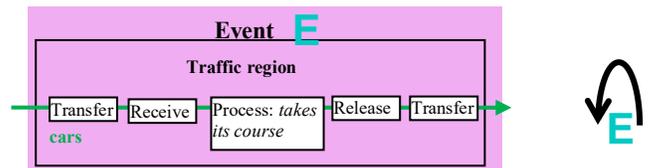

**Fig. 3** The event *Traffic as an existing thing*.

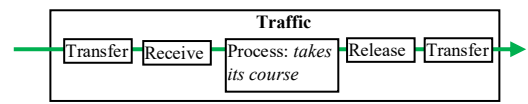

**Fig. 4** The region *Traffic as a subsisting thing*.

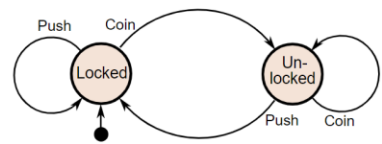

**Fig. 5** Turnstile state machine (from [19]).

Semantically, a *behavior* means whatever a system *does* that is publicly observable. It is a description of *doing* [22]. A *state* is a description of the *status* of a system that is waiting to execute a set of actions. It is a description of *having something done* [22].

The turnstile is a *system* that includes a passenger and a machine. The passenger does actions, and the machine has something done to it. So, the machine, not the system, does the waiting, and the passenger, not the system, does the actions. Therefore, it seems the state diagram involves missing implicit knowledge, a mixture of whom does what and ambiguity between doing something and having something done. By contrast, a TM contains only actions—in fact, only five actions—in any system: create, process, release, transfer and receive.

A static description is a sequence of states that a system passes through, like the model [17] discussed in the previous subsection, which is built using types such as *paint* and *dry* that are nonindividual (nonobject) notions. The occurrences of paint and dry are specified in another diagram.

Dynamic models, in contrast, describe how those states unfold in time [23]. They are used to understand how the system changes over *time*. In the turnstile example, Fig. 5 is built from *types* (not instances): coin, locked, push, etc. Therefore, it is a static representation. A static description is a sequence of states that a system passes through, like the model [17] discussed in the previous subsection, which is built using types such as *paint* and *dry* that are nonindividual (nonobject) notions. Dynamic models, in contrast, describe how those states unfold in time [23]. They are used to understand how the system changes over *time*.

**Static TM model**: Fig. 6 shows the static TM model. Coin is received to trigger (dashed arrow number 1) releasing the lock to the open position (2), pushing past (3) triggers moving the lock to the locked position (4).

**Dynamic TM model**: To construct this model, we need to define what an event is. A TM event is constructed from a sub-diagram of the static diagram (called *region*) and time. Fig. 7 shows the event *Coins are deposited in the turnstile machine*. For simplicity, we will represent the event by its region. Fig. 8 shows the dynamic system with four events, $E_1$–$E_4$. Fig. 9 shows the behavioral model of the turnstile system in terms of the sequence of events.

### D. Solution: A Single Multilevel Model

The previous section illustrated the idea of vertical modeling. The work in this article embraces modeling based on "everything is event" instead of "everything is object." The approach is called *occurrence-only* modeling, in which an *occurrence* means an *event* or *Process* where a Process (capitalized to distinguish it from "process" as an action in a TM) is defined as an assembly of events that form a whole (i.e., a high-level event) [15]). An event's presence (signifies an occurrence) is defined in time and "region" (a portion of the description of the potential world), and the event's absence (negation) is its region [15].

In a TM, *objects* are nothing more than long events. The underlining paradigm includes (a) Stoic ontology that has two types of being, *existence* and *subsistence*, (b) a TM that limits activities to five generic *actions* and (c) Lupascian logic, which handles *negative events*.

Our work in this paper focuses on TM modeling. We claim that a TM as a single multilevel model can potentially be a unifying diagrammatic language for almost all current modeling languages. To substantiate such a claim, we redraft in TM almost all diagrammatic representations in [4]. The redrafting diagrams as Fig. 10 shows. The results seem to indicate no difficulties in representing these diagrams in a TM. Note that Fig. 10 includes BPNM, ER and sequence diagrams referring to previous publications, in which these diagrams were represented in a TM.

### E. Paper's Structure

The next section comprises a general review of the occurrence-only framework of modeling that includes TM. Sections 3 and 4 consist of converting the following modeling tool diagrams to TM: textual requirements, context modeling, class diagram, use case, data flow diagram, activity diagram and state diagram.

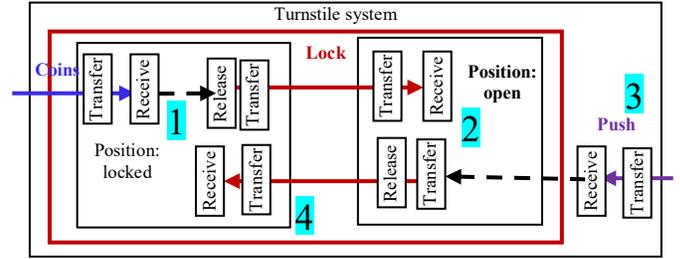

**Fig. 6** *TM static model of the turnstile system.*

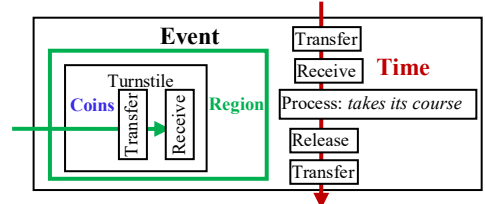

**Fig. 7** *The event Coins deposited in the turnstile machine.*

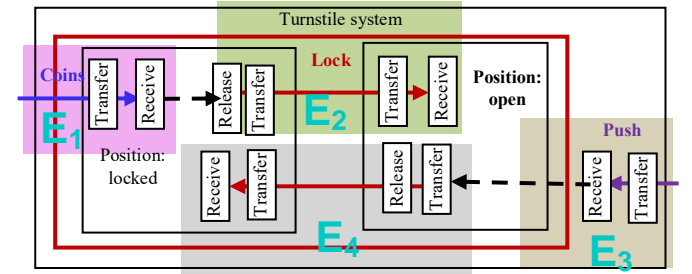

**Fig. 8** *TM dynamic model of the turnstile system.*

$$E_1 \rightarrow E_2 \rightarrow E_3 \rightarrow E_4$$

**Fig. 9** *TM behavioral model of the turnstile system.*

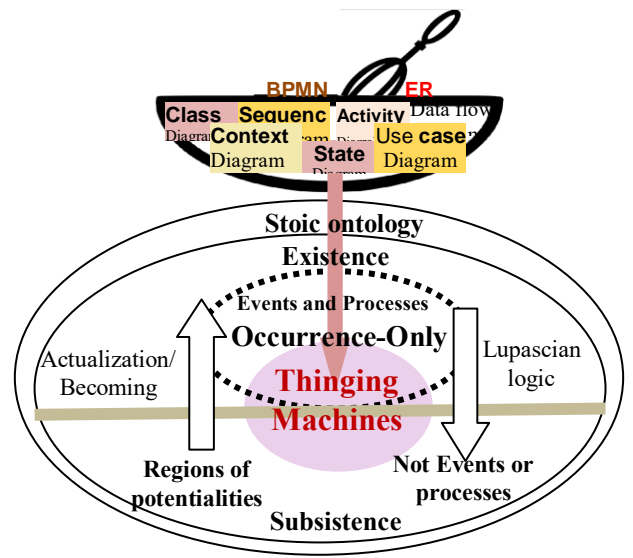

**Fig. 10** *A general framework of changing diagrams into a TM.*

## II. OCCURRENCE-ONLY MODELING

Occurrence-only conceptual modeling is founded on three grounds [15], TM, Stoic ontology and Lupascian logic. Here, we emphasize TM materials. More information on Stoic ontology and Lupascian logic can be found in [14]. The TM basic modeling entity is called a *thimac* (*thi*ng/*mac*hine) because it is conceptualized with the dual nature of a thing and machine. The machine consists of five actions: *create*, *process*, *release*, *transfer* and *receive*. (See Fig. 11). A thimac as a thing is created, processed, released, transferred and received. A thimac as a machine creates, processes, releases, transfers and receives. Therefore, instead of the famous saying "everything flows," in a TM, every thimac creates, processes, releases, transfers and/or receives, and every thimac is created, processed, released, transferred and/or received.

The TM machine actions are described as follows.
1) *Arrive:* A thing arrives to a machine.
2) *Accept:* A thing enters the machine. For simplification, we assume that arriving things are *accepted* (see Fig. 11); therefore, we can combine the *arrive* and *accept* stages into the *receive* stage.
3) *Release:* A thing is ready for transfer outside the machine.
4) *Process:* A thing is changed, handled and examined, but no new thing results.
5) *Transfer:* A thing is input into or output from a machine.
6) *Create:* A new thing manifested in a machine

Additionally, the TM model includes a *triggering* mechanism (denoted by a dashed arrow in this article's figures), which initiates a (non-sequential) flow from one machine to another. Moreover, each action may have its own memory storage (denoted by a cylinder in the TM diagram) of things. For simplicity, we may omit *create* in some diagrams because the box representing the thimac implies its beingness (in the model).

## III. STATIC MODELS

Starting with this section, we convert diagrams from [4] into TM diagrams discussed in the introduction. The aim is to substantiate our claim that TM as a single multilevel model can, potentially, be a unifying diagrammatic language for almost all current modeling languages.

According to [11], since the inception of the notion "Everything is an object," the unification principle has been the engine most helpful in the direction of simplicity, generality and power of integration. The idea of an object was a scheme unifying data and process. According to [4], to make the complexity of the modeling manageable, various views of the system are represented through diagrams. Each diagram is based on a specific diagram type, which in turn is defined via a modeling language. A number of diagram types and associated modeling languages are available for requirements modeling.

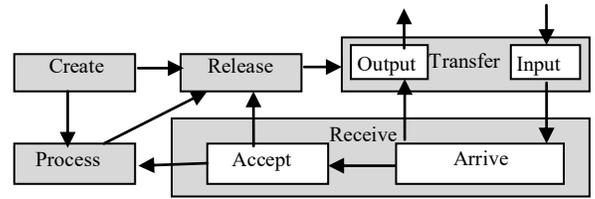

**Fig. 11 Thinging machine.**

The selection of the diagram type depends on the purpose, which thus determines which specific system requirements should be documented and which persons are the "target audience" for the requirements models. Diagram types used are those that allow the modeling of process-oriented aspects, such as event-based process chains or BPMN diagrams as part of the business analysis, as well as UML activity diagrams [4].

### A. Text and Diagram

Fig. 12 from [4] shows the difference between "textual" and "modeled" requirements and Fig. 13 shows its corresponding static TM model. In Fig. 12, the left panel shows textual requirements that specify necessary *behavior* in relation to the input of data via an entry screen. The right panel shows the requirements modeled in which "a model is regarded as an abstracting image of the properties of a system" [4].

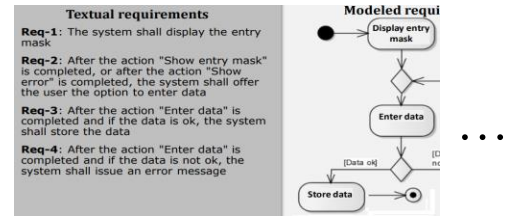

**Fig. 12 Example of requirements given by [4].**

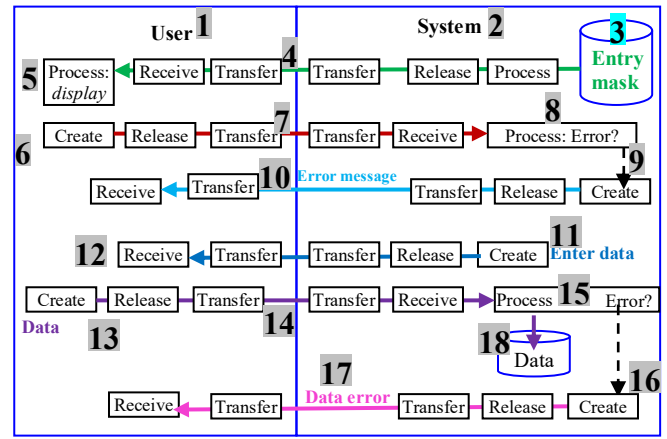

**Fig. 13 Static TM specification of the given requirements.**

This example shows that "the interactions of the various aspects of the required system behavior are explicitly visible in the modeled requirements, whereas this information is only implicitly present in the textual requirements" [4].

Fig. 13, the static TM diagram that corresponds to Fig. 12, involves the user (Number 1 in the figure) and the system (2). The entry mask (3) is retrieved and sent to the user (4) to be displayed (5). The user is supposed to input something (6) that flows to the system (7) where it is examined (8). If the input is wrong, then this triggers (9) an error message sent to the user (10). Otherwise, the user is asked (11) to enter data (12). The user enters the data (13) that moves to the system (14) to be examined (15). If there is an error in the data, then this creates a data error message (16) sent to the user (17); otherwise the data is stored (18).

Fig. 13 is an engineering schema based on the static specification and forms the foundation of the dynamic model. As before, for the sake of simplification, events will be represented by their regions. The following more meaningful events are more optimal than the generic events (events of the five generis actions) as shown in Fig. 14; thus, we list the following events.

$E_1$: The system sends the entry mask to be displayed to the user.
$E_2$: The user sends input to the system.
$E_3$: The system finds an error in the input.
$E_4$: An error message is sent to the user.
$E_5$: The system requests to enter data.
$E_6$: The user sends data.
$E_7$: The system finds an error in the data.
$E_8$: The system sends a data error message.
$E_9$: The system stores the data.

Fig. 15 shows the behavior diagram.

### B. Context Modeling

A challenge in requirements engineering is understanding the *context* of the system under development. A context diagram is used to identify the necessary interfaces between the system under development and its context. It is the part of a system's environment and relevant for communicating the system and its requirements [6].

Fig. 16 shows an example of a context diagram using structured analysis. The overall system in Fig. 16 (an early warning system in the mining industry) is represented as a circle in the middle. The human is labeled as a stick figure, and the organizational and technical neighboring systems are represented as boxes. The interface is modeled in the form of data flows to and from neighboring systems [4]. Fig. 17 shows the corresponding context diagram using TM modeling. We used a simplification where the actions, release, transfer and receive, are not shown.

### C. Class Diagram

According to [4], a "class" is a template that defines the common properties of many objects. The objects are then referred to as instances of these classes. The class diagram in Fig.18 includes a "part/whole" UML relationship. For the sake of simplicity, do not include *starting point* (see Fig. 18) in the corresponding model, Fig. 19.
6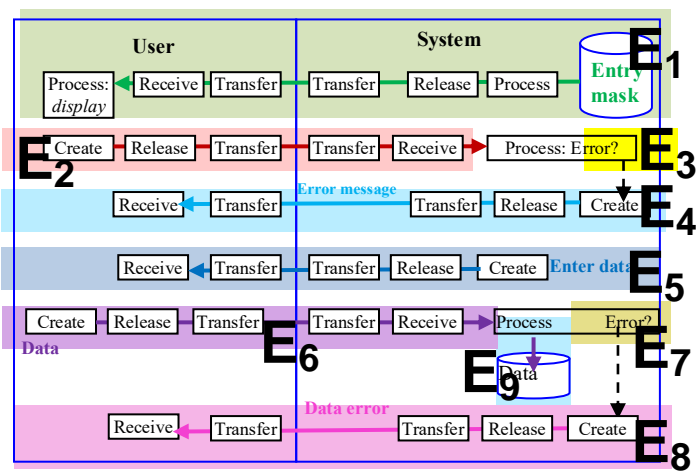

Fig. 14 The dynamic diagram

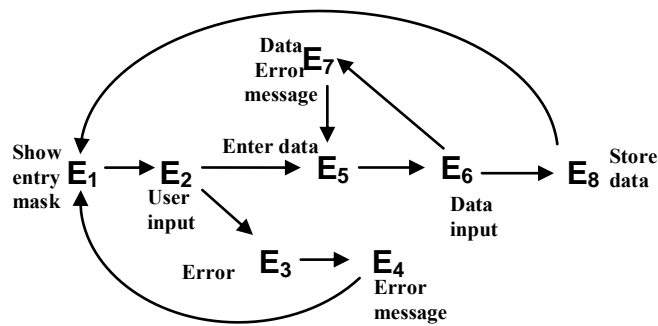

Fig. 15 The behavior diagram

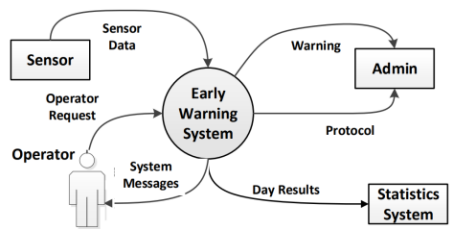

Fig. 16 Context diagram (from [4]).

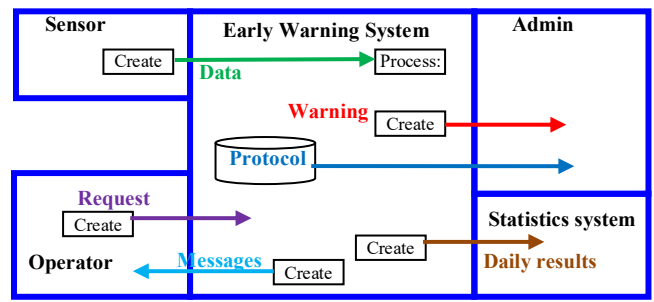

Fig. 17 Corresponding context diagram

*1) Static model*

Fig. 19 shows the static TM model, where the *route* contains the *destination* as part of its composition that cannot exist as a separate part, detached from a specific *route*. *Place of interest* in the figure involves the so-called aggregation with *route* where a *route* has zero (null value) or more *place of interest*, and a *place of interest* has at most one route (at a time).

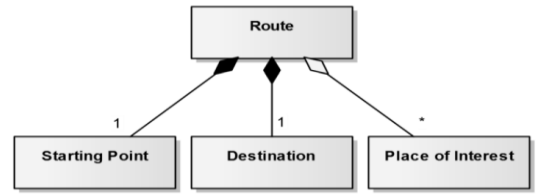

**Fig. 18 Sample class associations (from [4])**

The process of constructing the *route* starts at pink number **1**, and it is assumed that *route* **2** contains data: the ID **3** of the route, and other attributes **4** (not shown in the figure) and also *destination* **5** as a subthimac of the *route*. These fields form the composite part of route **6** that *places of interest* **7** will complete. Thus, the *destination* is part of the route that exists and perishes with the *route*.

*Place of interest* **7** is part of the *route* that is injected in route as follows. The *place of interest* is received **8** (independently from route) and processed **9** to extract the **place of interest** ID **10** and its *route* ID **11**. We assume that the *place of interest* data includes its route ID to be inserted in. This data is necessary to,

1) Insert the place of interest in its route **12** (the brown box labeled *Assigning Places to route*).

2) Enforce the constraints as follows.

**Enforcing that a** *place of interest* **is associated with only on route:** The route ID and *place of interest* ID **13** are processed **14** to create the tuple (ID, ID) **15** that flows to the system **16**. There, the tuple (ID, ID) is compared **17** with a similar tuple in a table **18** that registers all current existing relationships between IDs of routes and places of interests. Thus, if the given new destination already has different route, an error occurs **19**; otherwise, **20** the tuple (ID, ID) is added to the table if it is not already there. Thus, inserting the *places of interest* in the route is performed

Fig. 19 can easily be more elaborated (e.g., searching the (ID, ID) table), and it can easily be simplified.

*2) Dynamic model that corresponds to the class*

Fig. 20 shows the dynamic model. The dynamic model operates on the assumption that route data are stored in one file that includes multi-fields for places of interest.

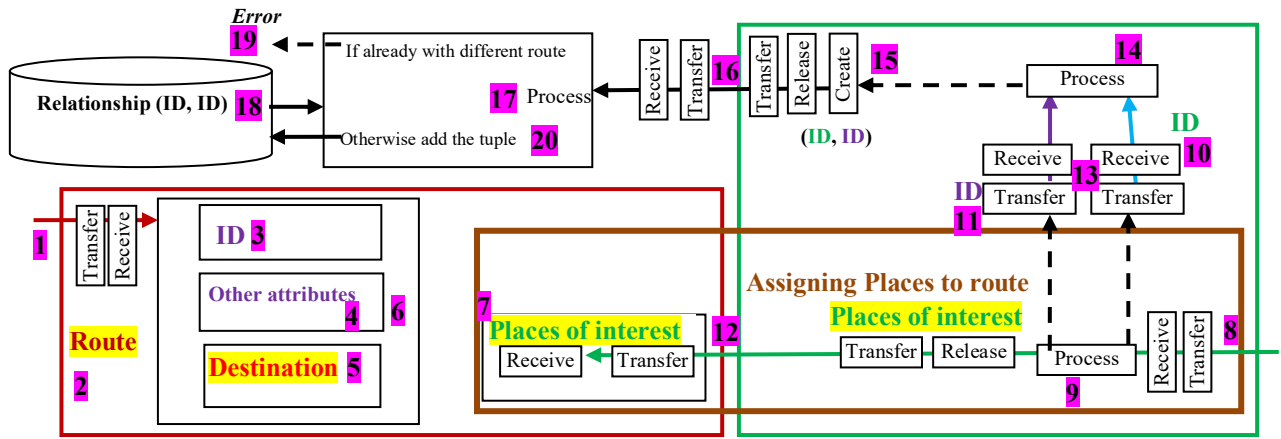

**Fig. 19 The static model of the class associations in the given example**

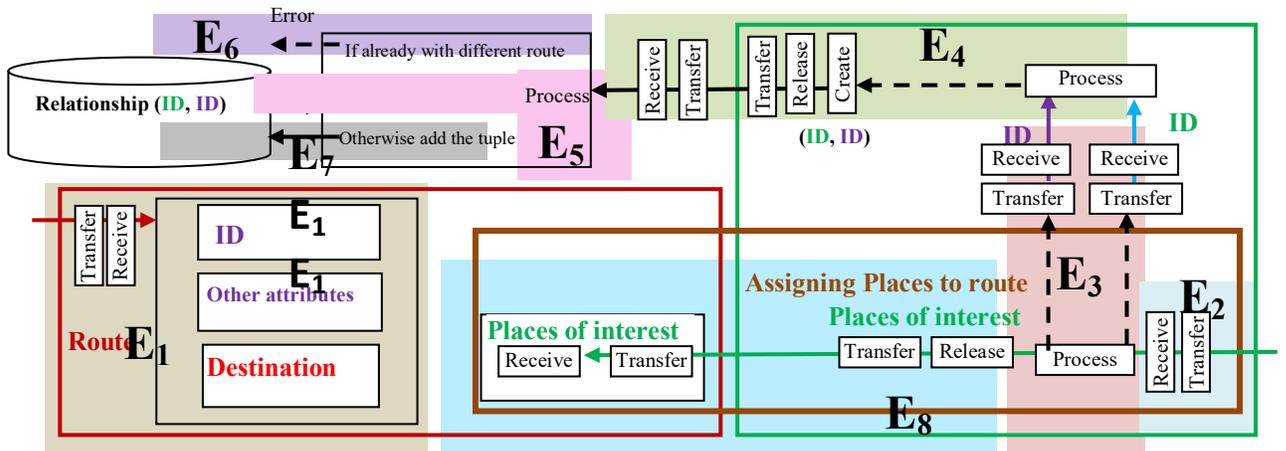

**Fig. 20 The dynamic model of the class associations in the given example.**

The dynamic model captures inputting one route with places of interest that is input one after another. The selected events are as follows:

$E_1$: Input route, including destination,
$E_2$: Input place of interest,
$E_3$: Extract the two IDs of a route and its place of interest,
$E_4$: Construct the IDs tuple
$E_5$: Search the ID tuples for whether the *place of interest* is already associated with the other *route*,
$E_6$: If the place of interest is already associated with a route, then issue an error message,
$E_7$: Add the IDs tuple to the table if not already there, and
$E_8$: Insert the place of interest in the route.

Fig. 21 shows the behavioral model for this Process.

## IV. DYNAMIC MODELING

According to [4], dynamic views "offer a lot of different abstraction criteria for specifying different aspects of the functionality." Ref. [4] provides the following types of dynamic views: *use case view*, *control flow-oriented view*, *data flow-oriented view*, *state-oriented view* and *scenario view*.

### A. Use Case

Fig. 22 shows a sample use case. Fig. 23 shows the corresponding TM model. Notably, the difference between the two representations is not just a matter of replacing the stickman with boxes. In TM modeling, the diagram in Fig. 23 is the *base* model that is enhanced repeatedly at the behavior level in contrast to the stickman notation that disappears in the next modeling phases.

### B. Data Flow Diagram

According to [4], data flow diagrams are often used to model requirements from a data flow-oriented perspective. Such a diagram can easily be constructed in different levels of TM models.

### C. Activity Diagram

According to [4], UML *activity diagrams* can model requirements from the control flow perspective, and these diagrams are useful mainly for communication between the persons involved: "the completeness of the specification can be achieved with supplementary activity descriptions."

Ref. [4] introduces a model for activity diagrams, "which are relevant for requirements engineering: the *interruptible* activity region" where the user terminates the activity by clicking on "Cancel." However, the example is not clear; hence, we illustrate the notion of *interruptible* activity from an example taken from [24] as shown in Fig. 24.

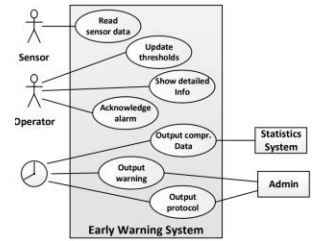

**Fig. 22 Sample use case (from [4]).**

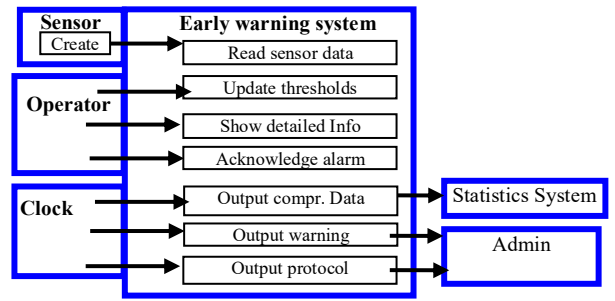

**Fig. 23 TM model that corresponds to the given use case diagram.**

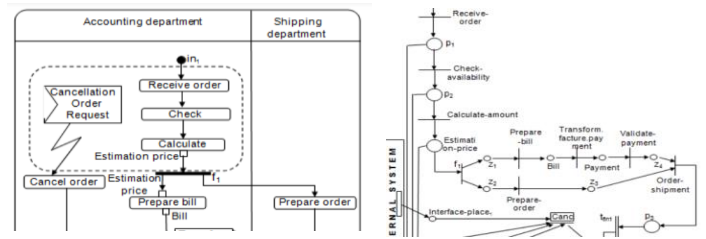

**Fig. 24 Activity diagram of the process *order commercial activity*, including interruptible region (left) and its Petri Nets cancellation event.**

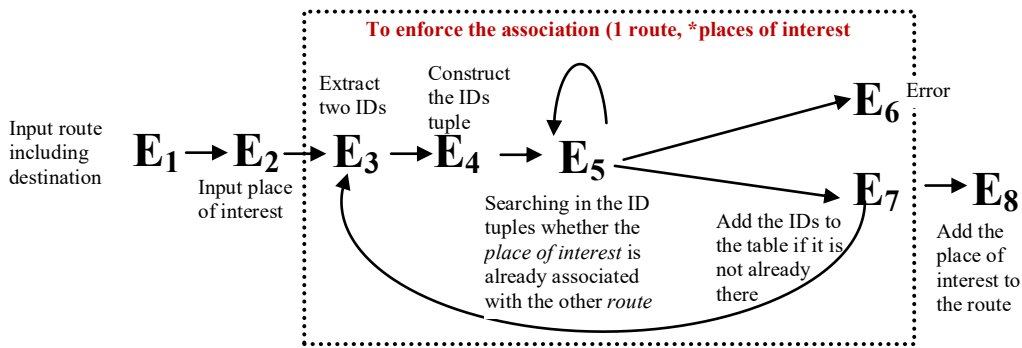

**Fig. 21 The behavioral model for constructing one route.**

The example involves a commercial order process in which a customer belonging to the system could intervene and stop the order processing procedure. Ref. [24] modeled this case using an activity diagram of the process order activity, including the interruptible region (left) and its corresponding Petri Nets cancellation event (right).

Fig. 25 shows the corresponding TM model. First, an order request arrives at the system (pink number 1). The request is processed to check the request (2). Assuming it is okay, the system creates the order (3) used to create an estimated price (4) that is sent to the customer (5). To save space in this paper, we will not describe the rest of the order procedure. Additionally, the customer can interrupt the order operations by a cancellation signal (6) that is processed (7) to cancel order progress at any phase.

Fig. 26 shows the dynamic of this ordering system. Fig. 27 shows the behavior system. Note that **$R_i$** in Fig. 27 represents the region of $E_i$, that is the **negative event**, "not (or stop) $R_i$" *reverting* to the static level as described in Section 2. In the behavior model, if the cancelation event $R_7$ coincides with any of the other events, then the event reverts to the static level, halting any progress in the order processing. Fig. 28 illustrates the progress of events in such a system.

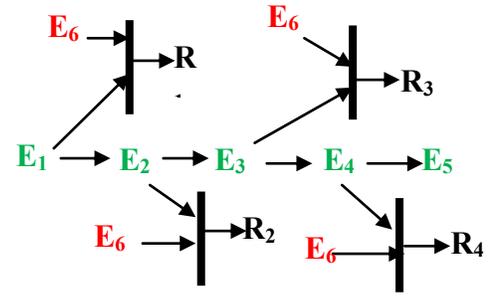

**Fig. 27 The behavior model of the order system.**

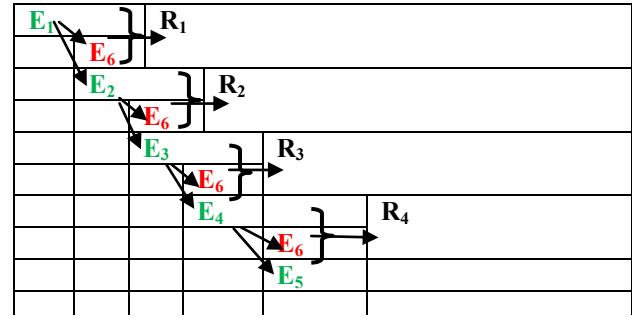

**Fig. 28 Illustration of progress of events**

### D. State Diagram

According to [4], the state space of the system is modeled in the state-oriented view within the dynamic view. The states and state changes observable at the interface between the system and the system context are modeled in this view. A state change of the system can be triggered by an event. Ref. [4] gives a typical example of the states of the object *request for leave* in a business-oriented system. This example does not have enough details to be understood completely by the author of this paper. However, the literature has many state diagrams of *request for leave* in business-oriented systems, and we selected an example (Fig. 29) to model taken from [25] to demonstrate our aim.

The state diagram in Fig. 29 involves the following: an employee submits the request, the supervisor approves the timesheet and the HR department reviews and finalizes the decision. The states are: drafting the request, pending manager's approval, returned, pending HR's approval, approved or rejected.

Fig. 30 shows the corresponding TM model. First, the employee submits the leave request (blue Number 1) and sends it to the manager (2).

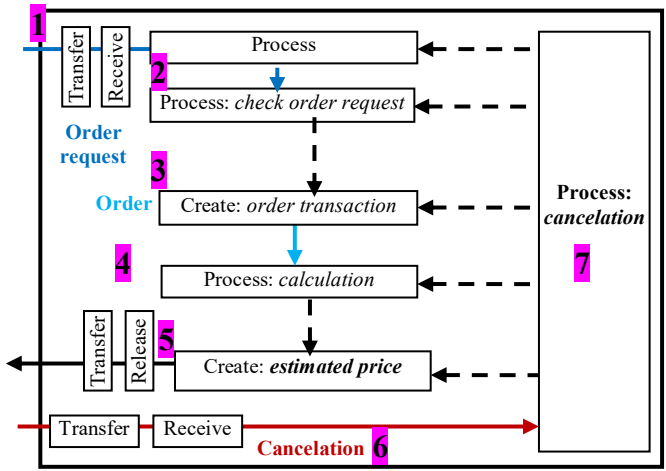

**Fig. 25 Static model of various phases of a customer's order**

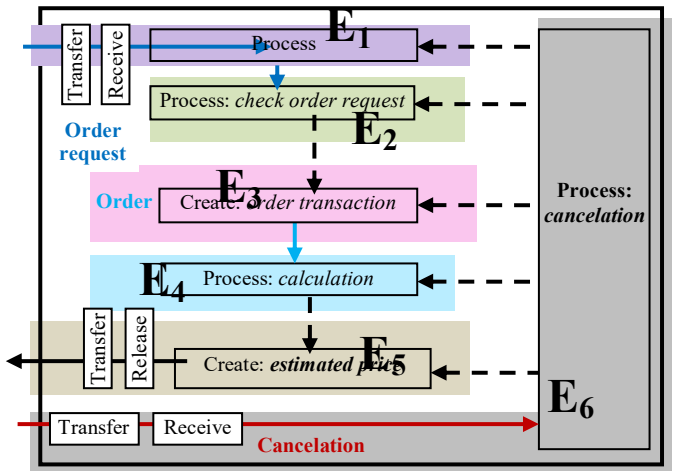

**Fig. 26 Dynamic model of the order system**

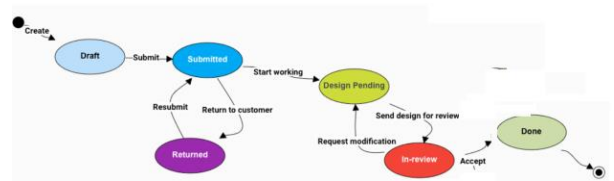

**Fig. 29 States of a *request for leave* (from [25]).**

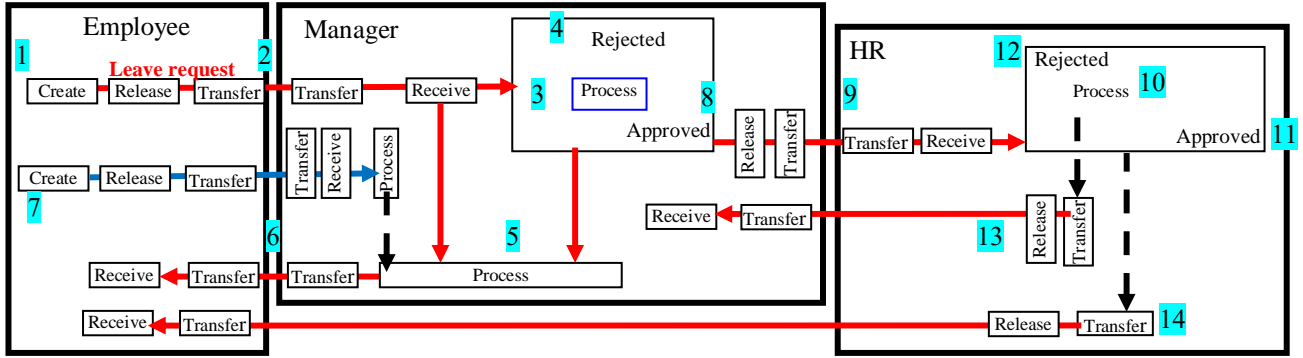

**Fig. 30 Static model**

The manager processes (3) the request to either reject (4) it or return it to the employee (5 and 6). The employee can also recall (7) the leave request, returning it to him/her. We assume that this recall proceeds processing the halt request. The manager may approve (8) the leave request and forward it to HR (9). HR processes (10) the leave request to approve it (11) or reject it (12) and returns it to the manager (13) or sends to the employee (14).

Fig. 31 and 32 show the dynamic and behavioral models, respectively.

## V. CONCLUSION

The importance of unification in modeling in requirements engineering hardly needs justification. This unified modeling enables systematizing the domain and minimizes internal inconsistencies. The dominant UML-based methodology aims at "unifying notation that incorporates the best of a number of other notations as well as current best practice in one generally applicable notation" [26]. This paper attempts to demonstrate another type of unification in which a unified model provides a coherent account of unconnected notions and reveals connections between phenomena believed to be unrelated.

In this paper, we focused on two approaches in this context: multimodeling that applies simultaneous usage of dissimilar notations vs. uni-modeling that utilizes a single framework of notations. Of course, the issue needs further investigation, but the TM model is clearly a viable approach to accomplishing the alternative type of modeling unification.

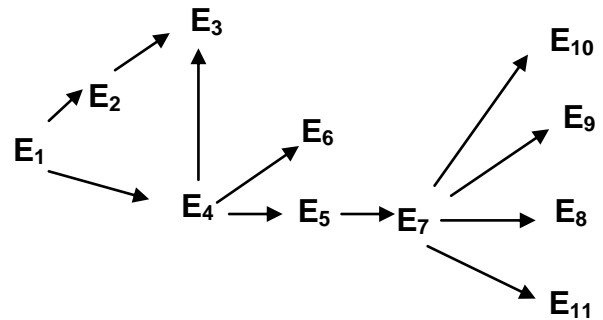

**Fig. 32 Behavioural model**

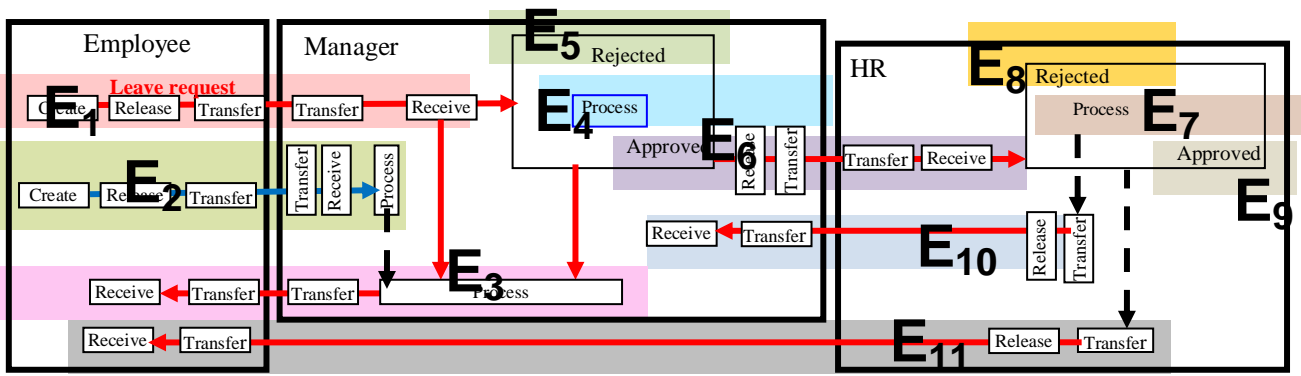

**Fig. 31 Dynamic model**